\documentclass[12pt]{article}

\usepackage{color}

\usepackage{ascmac}
\usepackage{amsmath}
\usepackage{amssymb}

\newcommand{\beq}{\begin{eqnarray*}}
\newcommand{\eeq}{\end{eqnarray*}}
\makeatletter
\renewcommand{\theequation}{\thesection.\arabic{equation}}
\@addtoreset{equation}{section}
\def\eqnarray{%
\stepcounter{equation}%
\let\@currentlabel=\theequation
\global\@eqnswtrue
\global\@eqcnt\z@
\tabskip\@centering
\let\\=\@eqncr
$$\halign to \displaywidth\bgroup\@eqnsel\hskip\@centering
$\displaystyle\tabskip\z@{##}$&\global\@eqcnt\@ne
\hfil$\displaystyle{{}##{}}$\hfil
&\global\@eqcnt\tw@$\displaystyle\tabskip\z@{##}$\hfil
\tabskip\@centering&\llap{##}\tabskip\z@\cr}
\makeatother
\newtheorem{theorem}{Theorem}[section]
\newtheorem{lemma}[theorem]{Lemma}

\newtheorem{remark}[theorem]{Remark}

\newsavebox{\toy}
\savebox{\toy}{\framebox[0.65em]{\rule{0cm}{1ex}}}

\def\nlni{\par\ifvmode\removelastskip\fi\vskip\baselineskip\noindent}

\begin{document}
\setlength{\baselineskip}{15pt}
\title{
The 
scaling limit of eigenfunctions for 1d random Schr\"odinger operator
}
\author{
Fumihiko Nakano
\thanks{
Department of Mathematics,
Gakushuin University,
1-5-1, Mejiro, Toshima-ku, Tokyo, 171-8588, Japan.
e-mail : 
fumihiko@math.gakushuin.ac.jp}
}
\maketitle
\begin{abstract}
We 
report our results on the scaling limit of the eigenvalues and the corresponding eigenfunctions for the 1-d random Schr\"odinger operator with random decaying potential. 
The formulation of the problem 
is based on the paper by Rifkind-Virag 
\cite{RV}. 
\end{abstract}


\section{Introduction}
In this note
we consider the following one-dimensional Schr\"odinger operator with random decaying potential : 
\beq
H := 
- \frac {d^2}{dt^2} + a(t) F(X_t) 
\eeq
where 
$a \in C^{\infty}({\bf R})$, 
$a(-t) = a(t)$, 
$a(t)$ 
is monotone decreasing for 
$t>0$ 
and 
\beq
a(t) = t^{- \alpha} (1 + o(1)), 
\quad
t \to \infty
\eeq
for some 
$\alpha > 0$.
$F \in C^{\infty}(M)$
is a smooth function on a torus 
$M$
such that 
\beq
\langle F \rangle
:=
\int_M F(x) dx = 0
\eeq
and 
$\{ X_t \}_{t \in {\bf R}}$
is the Brownian motion on 
$M$.
Since 
$a(t) F(X_t)$
is a compact perturbation with respect to 
$(-\triangle)$, 
the spectrum  
$\sigma (H) \cap (-\infty, 0)$
on the negative real axis is discrete.
The spectrum 
$\sigma (H) \cap [0, \infty)$ 
on the positive real axis is 
\cite{KU} : 
\beq
\sigma (H) \cap [0, \infty)
\mbox{ is }
\left\{
\begin{array}{lc}
\mbox{ a.c. } & (\alpha>1/2) \\
\mbox{ p.p. on }[0, E_c] \mbox{ and s.c. on }[E_c, \infty) 
& (\alpha=1/2) \\
\mbox{ p.p. } & (\alpha < 1/2) \\
\end{array}
\right.
\eeq
where 
$E_c$ 
is a deterministic constant.
For the 
level statistics problem, 
we consider the point process 
$\xi_{L, E_0}$ 
composed of the rescaling eigenvalues 
$\{ L(\sqrt{E_j (L)} - \sqrt{E_0}) \}_j$ 
of the finite box Dirichlet Hamiltonian 
$H_L := H |_{[0, L]}$
whose behavior as 
$L \to \infty$
is given by 
\cite{KN1, N2, KN2}
\beq
\xi_{L, E_0} 
\stackrel{d}{\to}
\left\{
\begin{array}{lc}
Clock(\theta(E_0)) & (\alpha>1/2) \\
Sine(\beta(E_0)) & (\alpha=1/2) \\
Poisson(d \lambda / \pi) & (\alpha<1/2)
\end{array}
\right.
\eeq
where 
$Clock (\theta) := 
\sum_{ n \in {\bf Z}} \delta_{n \pi + \theta}$, 
is the clock process for some random variable 
$\theta$
on 
$[0, \pi)$,
and 
$Sine(\beta)$ 
is the Sine$_{\beta}$-process which is the bulk scaling limit of the Gaussian beta emsemble \cite{VV}.
For 
$\alpha = 1/2$, 
$\beta (E_0) = \tau(E_0)^{-1}$ 
is equal to the reciprocal of the Lyapunov exponent 
$\tau(E_0)$ 
such that the solution to the Schr\"odinger equation 
$H \varphi = E \varphi$ 
has the power-law decay : 
$\varphi(x) \simeq |x|^{- \tau (E)}$, 
$|x| \to \infty$.
Since
$\lim_{E_0 \downarrow 0} \beta(E_0) = 0$ 
and 
$\lim_{E_0 \uparrow \infty} \beta (E_0) = \infty$, 
small (resp. large) 
$E_0$ 
corresponds to small (resp. large) repulsion of eigenvalues, which is consistent to the following fact \cite{AD, N3} : 
\beq
Sine (\beta) 
\stackrel{d}{\to} 
\left\{
\begin{array}{ll}
Poisson (d \lambda / \pi) & (\beta \downarrow 0) \\
Clock(unif[0, \pi)) & (\beta \uparrow \infty)
\end{array}
\right.
\eeq
In this note, 
we consider the scaling limit of the measure corresponding to the eigenfunction of 
$H_L$
under the formulation studied by Rifkind-Virag 
\cite{RV}. 
To formulate 
the problem, we need some notations. 
Let 
$\{ E_j (L) \}_j$ 
be the positive eigenvalues of 
$H_L$, 
and 
$\{ \psi^{(L)}_{E_j(L)} \}$ 
be the corresponding eigenfunctions. 
We consider the associated random probability measure
$\mu^{(L)}_{E_j(L)}$ 
on 
$[0,1]$.
%
\beq
\mu^{(L)}_{E_j(L)} (dt)
:=
C
\left(
| \psi^{(L)}_{E_j(L)}(Lt) |^2 
+
\frac {1}{E_j(L)}
\left| 
\frac {d}{dt}
\psi^{(L)}_{E_j(L)} (Lt)
\right|^2
\right) 
dt.
\eeq
Let 
$J := [a,b] (\subset (0, \infty))$ 
be an interval, 
${\cal E}^{(L)}_J := 
\{ E_j (L) \}_j \cap J$ 
be the eigenvalues of 
$H_L$ 
on 
$J$, 
and 
$E_J^{(L)}$ 
be the uniform distribution on 
${\cal E}^{(L)}_J$.
Our aim is 
to consider the large 
$L$ 
limit of the eigenvalue-eigenvector pairs : 
\beq
{\bf Q} : 
\left(
E_J^{(L)}, \mu_{E_J^{(L)}}^{(L)}
\right)
\;
\stackrel{d}{\to} 
\;
?
\eeq
For 
d-dimensional discrete random Schr\"odinger operator, 
if 
$J$ 
is in the localized region, we have
\cite{KiN, N1}
\beq
\left(
E_J^{(L)}, \mu_{E_J^{(L)}}^{(L)}
\right)
\stackrel{d}{\to}
\left(
E_J, \delta_{unif[0, 1]^d}
\right)
\eeq
where 
$E_J$ 
is the random variable obeying 
$\frac {1_J(E)}{N(J)} d N(E)$, 
where 
$dN$
is the density of states measure.
Rifkind-Virag \cite{RV} 
studied the 1-d discrete Schr\"odinger operator with critical decaying coupling constant, and obtained that  
the limit of 
$\mu^{(L)}_{E_J^{(L)}}$
is given by an exponential Brownian motion. 

\ \ 
To state our result, we need notations further.
Let 
$N(E) := \pi^{-1} \sqrt{E}$
be the integrated density of states, 
$N(J) := N(b) - N(a)$, 
and 
\beq
\tau(E)
:= 
\frac {1}{8E}
\int_M | \nabla (L + 2i \sqrt{E})^{-1} F|^2 dx. 
\eeq
where 
$L$ 
is the generator of 
$(X_t)$. 
Moreover, let 
$E_J$
be the random variable whose distribution is equal to 
$N(J)^{-1} 1_J (E) d N(E)$, 
let 
$U$ 
be the uniform distribution on 
$[0,1]$, 
and let 
${\cal Z}$ 
be the 2-sided Brownian motion, where 
$E_J$, $U$, and 
${\cal Z}$
are independent.
\begin{theorem}
\beq
&&
\left(
E_J^{(L)}, \mu^{(L)}_{E_j(L)}
\right)
\\
&&
\stackrel{d}{\to}
\left\{
\begin{array}{ll}
\left(
E_J, 1_{[0,1]}(t) dt 
\right) & (\alpha > 1/2) \\
\left(
E_J, 
\frac {
\exp
\Bigl(
2 {\cal Z}_{\tau(E_J)\log \frac tU} - 2 \tau(E_J) 
\left|
\log \frac tU 
\right|
\Bigr)
dt
}
{
\int_0^1
\exp
\Bigl(
2 {\cal Z}_{\tau(E_J)\log \frac sU} - 2 \tau(E_J) 
\left|
\log \frac sU 
\right|
\Bigr)
ds
}
\right)
& (\alpha=1/2) \\
\left(
E_J, \delta_{unif[0,1]}(dt)
\right)
& (\alpha < 1/2) 
\end{array}
\right.
\eeq
\end{theorem}
\begin{remark}
\mbox{}\\
(1)
In the result of 
Rifkind-Virag \cite{RV}, 
`` $\log \dfrac tU$" 
in the statement of Theorem 1.1 for 
$\alpha = 1/2$ 
is replaced by 
$| t - U |$. 
\\
(2)
For the continuum 
1-dimensional operator with decaying coupling constant, we have the similar result with Theorem 1.1 except that, as above, 
`` $\log \dfrac tU$" 
for 
$\alpha = 1/2$ 
is replaced by 
$| t - U |$. 
\end{remark}

When 
$\alpha < 1/2$, 
this result is the same as that in 
\cite{KiN, N1}, 
while for 
$\alpha > 1/2$ 
this result is natural. 
For 
$\alpha = 1/2$, 
this result implies that, the localization center 
$U$ 
of the eigenfunction 
$\psi$
is uniformly distributed and 
$\psi$
has the power law decay around 
$U$ 
with Brownian fluctuation. 
Since 
$\lim_{E \downarrow 0} \tau(E) = \infty$ 
and 
$\lim_{E \uparrow \infty} \tau (E) = 0$, 
$\psi$
is localized (resp. delocalized) for 
$E \downarrow 0$
(resp. $E \uparrow \infty$) 
which is consistent with the previous picture.
%
\section{Sketch of Proof}
For the proof, 
we mostly follow the strategy in \cite{KiN, N1, RV}, except for some technical points.
\subsection{Step 1 : renormalize the radial coordinate}
In what follows, 
we describe the solution 
$x_t$ 
to the equation 
$H x_t = \kappa^2 x_t$ 
in terms of the Pr\"ufer variales : 
\beq
\left(
\begin{array}{c}
x_t \\ x'_t /\kappa
\end{array}
\right)
=
r_t (\kappa)
\left(
\begin{array}{c}
\sin \theta_t(\kappa) \\ \cos \theta_t(\kappa)
\end{array}
\right)
\eeq
Introducing 
$\rho_t(\kappa)$ 
defined by   
$r_t (\kappa) := \exp ( \rho_t (\kappa) )$, 
we have
\beq
\rho_t(\kappa)
=
\frac {1}{2 \kappa}
Im 
\int_0^t 
e^{2i \theta_s(\kappa)} 
a(s) F(X_s) ds
\eeq
Let 
$\kappa_{\lambda} := \kappa_0 + \frac {\lambda}{n}$, 
$\kappa_0 := \sqrt{E_0}$, 
$\widetilde{\rho}^{(n)}_t(\kappa)
:=
\rho_{nt}(\kappa)
-
\langle F g_{\kappa} \rangle
\int_0^{n} a(s)^2 ds$, 
$g_{\kappa} := (L + 2i \kappa)^{-1} F$, 
$t \in [0,1]$.
We then have
\begin{lemma}
If 
$\alpha = 1/2$, 
then
\beq
&&
\widetilde{\rho}^{(n)}_t (\kappa_{\lambda})
\stackrel{d}{\to}
\widetilde{\rho}_t(\lambda), 
\quad
t \in [0,1], 
\mbox{ locally uniformly}
\\
&&
d 
\widetilde{\rho}^{(n)}_t (\kappa_{\lambda})
=
\frac {
\tau(\kappa_0^2)
}{t}
dt
+
\sqrt{
\frac {
\tau(\kappa_0^2)
}{t}
}
d B_t^{\lambda}, 
\quad
t > 0
\eeq
where 
$\{ B_t^{\lambda} \}$
is a family of Brownian motion.
\end{lemma}
%
\subsection{Step 2 : limit of the local version}
Let 
$\Xi^{(n)}$
be the local version of our problem : 
\beq
\Xi^{(n)}
:=
\sum_j
\delta_{
\Bigl(
n 
\bigl(
\sqrt{E_j(n)} - \sqrt{E_0}
\bigr), 
\mu_{E_j(n)}^{(n)}
\Bigr)
}
\eeq
It then
follows that 
\begin{lemma}
$\Xi^{(n)} \stackrel{d}{\to} \Xi$, 
where
\beq
\Xi
&=&
\left\{
\begin{array}{ll}
\sum_{j \in {\bf Z}}
\delta_{j \pi + \theta}
\otimes
\delta_{ 1_{[0,1]}(t) dt }
& (\alpha > 1/2) \\
\sum_{ \lambda : Sine_{\beta} }
\delta_{
\lambda
}
\otimes
\delta
\Bigl(
\frac { 
\exp ( 2 \widetilde{\rho}_t(\lambda) ) dt
}
{
\int_0^1
\exp ( 2 \widetilde{\rho}_s(\lambda) ) ds
}
\Bigr)
& (\alpha = 1/2) \\
\sum_{j \in {\bf Z}}
\delta_{P_j}
\otimes
\delta_{\widetilde{P}_j}, 
\quad
&
(\alpha < 1/2) 
\end{array}
\right.
\eeq
where 
$\{ P_j \} : Poisson (d \lambda /\pi)$, 
$\{ \widetilde{P}_j \} : Poisson (1_{[0,1]}(t) dt)$.
The intensity measure of 
$\Xi$ 
is given by 
\beq
&&
{\bf E} 
\left[
G(\lambda, \nu) 
d \Xi(\lambda, \nu)
\right]
\\
&=&
\frac {1}{\pi}
\left\{
\begin{array}{ll}
\int d \lambda
{\bf E}
\left[
G
\left(
\lambda, 
1_{[0,1]} (t) dt 
\right)
\right]
& 
(\alpha > 1/2) \\
\int d \lambda
{\bf E}
\left[
G
\left(
\lambda, 
\frac {
\exp
\Bigl(
2 {\cal Z}_{
\tau(E_0) \log \frac tU
}
-
2 \tau(E_0)
\log 
\left| \frac tU \right|
\Bigr)
dt
}
{
\int_0^1
\exp
\Bigl(
2 {\cal Z}_{
\tau(E_0) \log \frac sU
}
-
2 \tau(E_0)
\log 
\left| \frac sU \right|
\Bigr)
ds
}
\right)
\right]
&
(\alpha = 1/2)
\\
\int d \lambda
{\bf E}
\left[
G
\left(
\lambda, 
\delta_{ U }
\right)
\right]
& 
(\alpha > 1/2) 
\end{array}
\right.
\eeq
where 
$U := unif [0,1]$.
\end{lemma}
%
\subsection{Step 3 : averaging over the reference energy}
Following 
\cite{RV}, 
we introduce
\beq
g_1 (x)
&:=&
(1 - |x|) 1(|x| \le 1)
\\
G_L(E)
&:=&
\sum_{E_j(L) \in J}
g_1 
\left(
L 
\left(
\sqrt{E_j(L)} - \sqrt{E_0}
\right)
\right)
\cdot
g_2 \left(
E_j(L), \mu^{(L)}_{E_j(L)}
\right)
\eeq
where 
$g_2 \in C_b ({\bf R} \times {\cal P}(0,1))$.
We compute 
$\int \dfrac {dN(E)}{N(J)} {\bf E}[ G_L (E) ]$
by the following two ways, and then equate them by the Fubini theorem, which leads to the conclusion. 
\\
(1)
Since 
$\int \frac {dN(E)}{N(J)} g_1 = 1 / (L \pi)$, 
we have
\beq
&&
{\bf E}\left[ 
\int \dfrac {dN(E)}{N(J)} G_L (E) 
\right]
\\
&=&
\frac {1}{N(J)}
\frac {1}{\pi L}
{\bf E}
\left[
\sum_{E_j(L) \in J}
g_2 \left(
E_j(L), \mu^{(L)}_{E_j(L)}
\right)
\right]
\\
&=&
{\bf E}
\left[
\frac {1}{
\sharp \{
\mbox{ eigenvalues of 
$H_L$ 
on 
$J$ 
}
\}
(1 + o(1)) 
}
\cdot
\frac {1}{\pi}
\cdot
\sum_{E_j(L) \in J}
g_2 \left(
E_j(L), \mu^{(L)}_{E_j(L)}
\right)
\right]
\eeq
(2)
\beq
&&
\int \dfrac {dN(E)}{N(J)} {\bf E}[ G_L (E) ]
\\
&=&
\int \dfrac {dN(E)}{N(J)} {\bf E}
\left[
\sum_{E_j(L) \in J}
g_1 
\left(
L 
\left(
\sqrt{E_j(L)} - \sqrt{E_0}
\right)
\right)
\cdot
g_2 \left(
E_j(L), \mu^{(L)}_{E_j(L)}
\right)
\right]
\\
& \sim &
\int \dfrac {dN(E)}{N(J)} {\bf E}
\left[
\int 
g_1 (\lambda)
g_2 (E, \mu)
d \Xi^{(L)} (\lambda, \mu)
\right]
\\
& \to &
\int \dfrac {dN(E)}{N(J)} {\bf E}
\left[
\int g_1 (\lambda) 
g_2 (E, \mu)
d \Xi(\lambda, \mu)
\right]
\\
&=&
\int \dfrac {dN(E)}{N(J)}
\frac {1}{\pi}
\left\{
\begin{array}{ll}
\int d \lambda
{\bf E}
\left[
g_2
\left(
E, 
1_{[0,1]} (t) dt 
\right)
\right]
& 
(\alpha > 1/2) \\
\int d \lambda
{\bf E}
\left[
g_2
\left(
E, 
\frac {
\exp
\Bigl(
2 {\cal Z}_{
\tau(E_0) \log \frac tU
}
-
2 \tau(E_0)
\log 
\left| \frac tU \right|
\Bigr)
dt
}
{
\int_0^1
\exp
\Bigl(
2 {\cal Z}_{
\tau(E_0) \log \frac sU
}
-
2 \tau(E_0)
\log 
\left| \frac sU \right|
\Bigr)
ds
}
\right)
\right]
&
(\alpha = 1/2)
\\
\int d \lambda
{\bf E}
\left[
g_2
\left(
E, 
\delta_{ U }
\right)
\right]
& 
(\alpha > 1/2) 
\end{array}
\right.\eeq

\vspace*{2em}

This work is partially supported by 
JSPS KAKENHI Grant 
Number .26400145(F.N.)

%

\end{document}